\documentclass[12pt]{article}

\usepackage{amsmath}
\usepackage{cite}
\usepackage{epsfig}

\numberwithin{equation}{section}

\def\Im{\mathop{\rm Im}}
\def\Tr{\mathop{\rm Tr}}
\newdimen\picraise
\newcommand\picbox[1]
{
  \setbox0=\hbox{\input{#1}}
  \picraise=-0.5\ht0 
  \advance\picraise by 0.5\dp0
  \advance\picraise by 3pt      
  \hbox{\raise\picraise \box0}
}

\begin{document}
\begin{tabbing}
\` Bonn TK 98-09 
\end{tabbing}
\baselineskip 10mm
\renewcommand{\thefootnote}{\fnsymbol{footnote}}
\begin{center}{\Large\bf 
The theory of quark confinement\footnote{   
The work was completed in Bonn under the Humboldt Research Award
Program.}\footnote{ 
The text was prepared for publication by  Yu.~Dokshitzer, C.~Ewerz, 
and J.~Nyiri, 
with help from J.~Bartels, A.~Kaidalov, A.~Mueller and A.~Vainshtein.
}}
\end{center}
\begin{center}{\large 
V.\,N.\ Gribov
}\end{center}
\baselineskip 7mm
\begin{center}{\it 
Landau Institute for Theoretical Physics, Moscow \\
and \\
Research Institute for Particle and Nuclear Physics, Budapest \\
and \\
Institut f\"ur Theoretische Kernphysik der Universit\"at Bonn 
}
\end{center}

\begin{abstract}
  This is the second of the two last papers by V.\,N.\ Gribov concluding
  his 20 year long study of the problem of quark confinement in QCD.
  In this paper the analytic structure of quark and gluon Green's
  functions is investigated in the framework of the theory of confinement
  based on the phenomenon of supercritical binding of light quarks.
  The problem of unitarity in a confining theory is discussed.  The
  write-up remained unfinished and so it is presented here.  The
  author was planning to emphasise the link between the electro-weak
  and strong interactions, and in particular the r\^ole of pions
  (Goldstone bosons) in confinement, to present an explicit solution
  for bound states, and to write down an analytic model for quark and
  gluon Green's functions corresponding to confinement.
\end{abstract}

\renewcommand{\thefootnote}{\arabic{footnote}}
\setcounter{footnote}{0}
%
\tiny
\normalsize

\section{Introduction
{\protect
\footnote{The author intended to rewrite this Introduction. 
In the present form it corresponds to the Introduction to the Lecture given 
by V.\,N.~Gribov 
at the 34$^{th}$ International School of Subnuclear Physics in Erice, Italy, 
in 1996 [7]. 
}}}

Almost ten years ago [1] I proposed a hypothesis according to
which quarks and gluons are confined due to the existence of light
quarks ($u$ and $d$) with Compton wavelengths much larger than the
radius of strong interaction defined by $\lambda_{\mbox{\scriptsize QCD}}$. 
The essence of 
this hypothesis is the supercritical phenomenon which is well-known in
QED where it is the following. If there exists a heavy nucleus of radius
$R$ with charge $Z$ exceeding a critical value $Z_{cr}$ (which is of
the order of 137), the vacuum of the light charged fermions
(electrons) becomes unstable due to the process of ($e^-,e^+$) pair
creation. The negative component of the pair, the electron, falls
into the heavy charge and the positive component goes to infinity.
The condition for this phenomenon to occur is that the Compton wave
length $1/m$ of the electron has to be much larger than the radius
of the heavy charge:
\begin{equation}
\label{1}
\alpha\sqrt{Z^2-Z_{cr}^2}\ln(mR)>\pi\> .
\end{equation}
The electron which falls into the centre forms together with the
source $Z$ an ion of charge $Z_1=Z-1$. If $Z-1>Z_{cr}$, this process
will continue until the vacuum becomes stable i.e.\ when
$Z_n=Z-n$ becomes less than the critical charge.

From the point of view of the Dirac equation the falling electron
has negative energy. On the other hand, the supercritical ion
with $Z_n<Z_{cr}$ is stable because of the Pauli principle. Indeed,
the electron cannot leave the ion, since all negative energy states
outside it are occupied.

In QCD this phenomenon can occur since, due to asymptotic freedom,
the colour charge can reach the critical value for any object.
Let us suppose that two heavy quarks are created with opposite
colours.
\begin{figure}[ht]
\begin{center}
\input{fig1.pstex_t}    
\end{center}
\caption{}
\label{Fig1}
\end{figure}
The gluonic vacuum polarisation increases the colour charges
of the quarks (see Fig.\ 1). When these colour charges become large enough, 
light antiquarks start to fall onto the heavy quark. However, the bound
state formed by a light antiquark and a heavy quark will be very
different from the bound states which appeared in QED. In QED we
had an ion with a charge $Z-1$. In QCD the corresponding state can
be colourless. The difference is due to the fact, that while
the large colour charge of the heavy quark results from gluonic
vacuum polarisation, the bound state is formed by two particles the
total charge of which equals zero, and this charge cannot be changed
by vacuum polarisation. This phenomenon can be called quantum
screening. Formally it means that the heavy quark $q_h$ is unstable and
has to decay into a meson and a light quark $q_l$
\begin{equation}
\label{2}
 q_h \rightarrow M + q_l .
\end{equation}
This case --- the confinement of the heavy quark --- is relatively
simple and almost independent of the properties of the light
quark~[2]. 
The problem of the light quark is more difficult. At first
sight it is not clear whether the "falling into the centre"
and the formation of a supercritical state can occur in the
system of a light quark and a light antiquark. It was shown [3],
however, that indeed the critical phenomenon exists in the light
$q\bar{q}$ system. Moreover, the critical coupling constant is
proven to be sufficiently small:
\begin{equation}
\label{3}
 \frac{\alpha_{cr}}{\pi} = \left( 1 - \sqrt{\frac{2}{3}}\, \right)
 \cdot \frac{3}{4} \>\simeq\> 0.14\>.
\end{equation}
There is an even more complicated question: what kind of
states can be formed as a result of this phenomenon? If the
appearing new state is a normal meson, the "falling into the centre"
leads to
\begin{equation}
\label{4}
q_l \rightarrow M + q_l
\end{equation}
instead of (\ref{2}). Equation (\ref{4}) can be satisfied if
there are not only positive energy quarks $q_+$ but also negative
energy quarks $q_-$. This result, although it looks very strange,
is not unexpected. Indeed, we already learned in QED that particles
which fall into the centre correspond to the negative energy solution
of the Dirac equation.

To resolve the problem of the negative energy states, Dirac supposed
that they are all occupied and therefore not observable. The
necessity to consider now both the positive and negative solutions
of the Dirac equation means only that Dirac's hypothesis is not
always true. If the interaction is strong enough, there is another
possibility: negative energy states might be only partly occupied
and positive energy states only partly empty. This is, however,
not in contradiction with the absence of negative energy particles in
the real world if stable supercritical bound states (mesons) exist
since in this case both the positive and negative energy quarks
will be unstable:
\[ q_+ \rightarrow M + q_- \]
\begin{equation}
\label{5}
q_- \rightarrow M + q_-q_-\bar{q}_-
\end{equation}
and, consequently, unobservable.

From the point of view of the fermionic spectrum in the vacuum,
(\ref{5}) means the following. In the case of weak interactions,
the fermionic spectrum has a structure corresponding to Fig.~\ref{Fig2}.
\begin{figure}[ht]
\begin{center}
\input{fig2.pstex_t}    
\end{center}
\caption{}
\label{Fig2}
\end{figure}
All levels above $q_0=\sqrt{\vec{q}\,^2+m^2}$ are empty, and the levels
below $q_0=-\sqrt{\vec{q}\,^2+m^2}$ are occupied. If the interaction
is 
stronger
than critical, quark-antiquark pairs are
present in the vacuum, and the spectrum has a structure which
roughly corresponds to Fig.~\ref{Fig3}: all levels are occupied in the
shaded regions and empty in the unshaded ones.
\begin{figure}[ht]
\begin{center}
\input{fig3.pstex_t}    
\end{center}
\caption{}
\label{Fig3}
\end{figure}
In terms of condensed matter physics, the curves
$q_0=\pm\sqrt{\vec{q}\,^2+m_f^2}$ correspond to the Fermi surface.
The change in the number of constituents in the theory
is also natural from the point of view of the physics of
condensed matter.
In a theory with weak coupling we had two constituents $q$ and
$\bar{q}$; a theory with supercritical coupling contains four
states $q_+$, $q_-$, $\bar{q}_+$ and $\bar{q}_-$. In non-relativistic
physics there are only particles (electrons) and
no antiparticles. But in a conductor we have particles and holes,
and holes are negative energy states (counting
the energy of the holes from zero and not from the Fermi energy
as it is usually done).
 
In a relativistic theory it is impossible to add a constant to the
energy of particles, and therefore we are forced to talk about
negative energy states which exist only inside our matter (vacuum).

This unusual quark spectrum has been discussed for several years.
Still, I have not been able to formulate a constructive theory because
I have not understood what meson had to be introduced as a
supercritical bound state. From the structure of the spectrum
shown in 
Fig.~\ref{Fig3} 
it is clear that there have to be different types
of excitations corresponding to different types of mesons. It is
natural to identify hole--particle excitations of the type 1 near
the Fermi surface with mesons like $\rho$, $\omega$ etc. 
Hole--particle excitations near the light cone can be identified with
the $a_0$ and $f_0$ mesons [4] and also with $\eta'$. But all these
mesons do not look like natural candidates for the lowest supercritical
bound states. Only recently did I recognise that there are strong
reasons to believe that the pseudoscalar octet, and the $\pi$-meson
in particular, are in fact the lowest supercritical bound states.
Most of these reasons are connected with the discussion of the nature
of the $\pi$-meson state which I presented in~[5].

As I have said before, the supercritical atom in QED is not a usual
Bohr-type atom, containing a definite number of electrons. Rather,
it is a collective state in which only the electron density is
localised around the heavy charge, and only at large distances
(much larger than the atomic radius) it looks like a state with
a definite electron number. A quasi-Goldstone state like a 
$\pi$-meson with a finite mass has the same properties. At short distances
(distances less than $\lambda_{\mbox{\scriptsize QCD}}$) it can be considered as a
collective state of the type 
$q_l \bar{q}_l - q_r \bar{q}_r$ (divergence of the pseudovector current;
$q_l$ and $q_r$ stand for left-handed and right-handed
quarks), whereas at large distances it is the two-particle state
$q_l\bar{q}_r - q_r \bar{q}_l$. Accepting this identification
with the $\pi$-meson as the lowest supercritical
state by introducing it explicitly in the equation for the
quark Green function (which will be discussed in the next
part of the lecture), it proves possible to find a solution
for this Green function which has properties corresponding to the
spectrum in
Fig.~\ref{Fig3}. 
This solution has two complex poles as functions
of energy in the complex energy plane in accordance with the two
types of quarks (constituent and current) we have discussed. It
has no pole on the real axis which guarantees that quarks as
propagating states do not exist. It has a soft singularity when
$q^2 \rightarrow 0$, reflecting the fact that quark currents exist
in the region where the hadrons are created. It is important to 
stress that in all these considerations I assume the coupling
constant $\alpha$ is saturated at a value not much larger than
$\alpha_{cr}$. Quark masses corresponding to the positions of the
two complex poles $m_{\pm}$ are of the order of
  \[ 
m_{\pm} \sim  \lambda_{\mbox{\scriptsize QCD}}\> 
\exp\left(-\frac{C_{\pm}}{\sqrt{\alpha-\alpha_{cr}}}\right).  
\]
The solution for the Green's function is self-consistent if the 
$\pi$-meson mass is close to $m_-$. This means that at least in the case
when $\alpha$ is not very large, the $\pi$-meson mass is defined
by strong interaction dynamics. This result contradicts the usual
point of view according to which the $\pi$-meson mass squared is
the product of the bare quark mass $m_0$ defined by weak interactions
and the condensate density $\langle\bar{\Psi}\Psi\rangle/f^2_{\pi}$
defined by strong interactions.

In what follows I will explain what type of equations have to be
written and solved in order to come to the conclusion I have
stated and to calculate quantities which have not been analysed
until now. For example, quark-gluon vertices and the gluon Green's
function have not been calculated yet. It is clear, however, that
the gluon Green's function we obtain will also have a
complex singularity due to the gluon decay into a $q\bar{q}$ pair.
Consequently, the gluon will also be confined. Equations for
hadronic amplitudes can also be written constructively if $\alpha$
is not very large. It can be shown that if the quark Green's
function has properties as described above, the hadronic
amplitude will have no singularities connected with
intermediate $q\bar{q}$ states, but will have singularities related
to the $\pi$-meson thresholds. The fact that the $\pi$-meson is
included intrinsically in the equation for the quark Green's
function and has a mass of the order of $m_-$ makes this statement
much less mysterious than it would look without it.

\section{The structure of the confined solution for the \\
Green's function of massless quarks}\label{II}

In the paper [6] we considered the solution for the quark
Green's function corresponding to the chiral symmetry breaking. We
discussed the importance of including the contribution of the
Goldstone boson in the equation. In the present paper we show that this
solution does not necessarily survive in the presence of the Goldstone
contribution which, essentially, reflects the softness of the
condensate and leads to the existence of a solution which has no
poles and which corresponds to confined quarks.

We will accept the equation for the quark Green's function in the
form
\begin{equation}
\label{2.1}
\partial^2 G^{-1}(q) = g(q)\partial_{\mu}G^{-1}(q)G(q)\partial_{\mu}
 G^{-1}(q) - \frac{3}{16\pi^2 f_{\pi}^2}\{i\gamma_5,G^{-1}(q)\}
 G\{i\gamma_5,G^{-1}(q)\}
\end{equation}
where $g(q)=\frac{4}{3}\frac{\alpha}{\pi}$ is supposed to behave
like in [6] (see Fig.~\ref{Fig4})  
\begin{figure}[ht]
\begin{center}
\input{fig4.pstex_t}    
\end{center}
\caption{}
\label{Fig4}
\end{figure}
and $f_{\pi}$ is the amplitude for the pion -- axial current transition,
\begin{equation}
\begin{split}
\label{2.2}
f^2_{\pi} 
  &=  \frac{1}{8}\int\frac{d^4 q}{(2\pi)^4i} \Tr\{i\gamma_5,G^{-1}\}
 G\{i\gamma_5,G^{-1}\}G
\>A_{\mu}A_{\mu} \\
  & \hspace{4mm} 
  +  \frac{1}{64 \pi^2 f_{\pi}^2} \int \frac{d^4 q}{(2\pi)^4 i}
 \Tr \big(\{\gamma_5,G^{-1}\} G\big)^4 .
\end{split}
\end{equation}
Before turning to the formal solution, let us discuss
its general properties and the difference in the structure caused
by the inclusion of the pion contribution.

Writing $G^{-1}(q)$ as
\begin{equation}
\label{2.3}
G^{-1}(q) = Z^{-1}(q)\big( m(q)-\hat{q}\big) \,,
\end{equation}
we obtain (\ref{2.1}) in the form
\begin{equation}
\label{2.4}
\partial^2 G^{-1} = g(q) \partial_{\mu}G^{-1}G\partial_{\mu}G^{-1} +
 \frac{m^2}{f^2} \frac{m-\hat{q}}{m^2-q^2}Z^{-1} , 
\end{equation}
where $f^2 = \frac{4}{3}\pi^2 f_{\pi}^2 $.
For the solution of the equation for $G^{-1}$ without pion contribution
(i.e.\ without the last term in (\ref{2.4})) the behaviour of $Z(q^2)$ and
$m(q^2)$ as functions of $q^2$ in the Euclidean region of negative
$q^2=-Q^2$ 
is shown in Fig.~\ref{Fig5}. 
\begin{figure}[ht]
\begin{center}
\input{fig5.pstex_t}    
\end{center}
\caption{}
\label{Fig5}
\end{figure}

\noindent
Here the dashed curve corresponds to the massive quark, $m_0\neq 0$, and
the solid curve to the massless quark, respectively. 
In the latter case
\begin{equation}
\label{2.5}
m(q^2) = -\frac{\nu^3}{Q^2} \>,\quad Q^2\to \infty \,.
\end{equation}
If we include the last term of (\ref{2.4}), the solution corresponding
to massive quarks disappears, as it was expected. The massless quark
solution (\ref{2.5}) will, however, survive because in this case the last
term in (\ref{2.4}) which can be called $\Delta_{\pi}$ is small at large
$Q^2$ values:
\begin{equation}
\label{2.6}
\Delta_{\pi} = - \frac{\nu^6 \hat{q}}{f^2 Q^6}Z^{-1}\>, 
\quad  Q^2\to \infty \,.
\end{equation}
Let us see what would be the effect of the pion contribution at small
$Q^2$. 
For $Q \ll m_c
=m(0)$ (\ref{2.4}) leads to the
following two equations:
\begin{equation}
\label{2.7}
\partial^2(Z^{-1}m) = \left(\frac{4g}{m_c^2} + \frac{1}{f^2}\right)
Z^{-1}m ,
\end{equation}
\begin{equation}
\label{2.8}
\partial^2 Z^{-1} + \frac{2}{q^2}q_{\mu}\partial_{\mu}Z^{-1} =
 \left(\frac{2g}{m_c^2} + \frac{1}{f^2}\right)Z^{-1} .
\end{equation}
We see that the pion produces a hundred percent correction to the
effective coupling because $f\sim m_c$ and $g\simeq 0.2$ if we are close
to the critical value of $g$. This change can be crucial.
We shall analyse the equation (\ref{2.6}) more carefully later. For
the time being, in order to see what can happen, let us accept (\ref{2.7})
and (\ref{2.8}) literally. These simple equations are easy to solve.
The solutions are
\begin{eqnarray}
\label{2.9}
Z^{-1}m &=& \frac{\mu_1^2}{Q}\, Z_1(Q/\mu_1) \>, \quad 
\frac{4g}{m_c^2} + \frac{1}{f^2} =\frac1{\mu_1^2} \>; \\
\label{2.10}
Z^{-1} &= & \frac{\mu_2^2}{Q^2}\,Z_2(Q/\mu_2) \>, \quad 
\frac{2g}{m_c^2} + \frac{1}{f^2} = \frac1{\mu_2^2}\>,
\end{eqnarray}
where $Z_1$ and $Z_2$ are 
solutions $Z_\nu$ 
of the Bessel equation with index $\nu=1,2$,  respectively. 
The concrete forms of $Z_1$ and
$Z_2$ depend on the boundary conditions imposed on the solutions. If we
want to preserve the behaviour of $Z^{-1}$ and $m$ at small $Q^2$ values,
we have to choose
\begin{equation}
\label{2.11}
Z_1  \propto  J_1(Q/\mu_1)\> , \qquad         
Z_2  \propto J_2(Q/\mu_2)\>,
\end{equation}
with $J_1$ and $J_2$ the Bessel functions. 
In this case, however,
the behaviour of $Z^{-1}$ and $m$ will change at large $Q^2$ values and
it is not obvious whether it will be possible to preserve the asymptotic
behaviour (Fig.~\ref{Fig5}) corresponding to asymptotic freedom. If our aim
is to keep the behaviour at large $Q^2$ unchanged we have to choose
$Z_1$ and $Z_2$ as superpositions of the singular and non-singular
solutions of the Bessel equation,
\begin{subequations}
\label{2.12}
\begin{eqnarray}
Z_1 & = & a_1 Y_1(Q/\mu_1) + b_1 J_1(Q/\mu_1)\> ,   \\
Z_2 & = & a_2 Y_2(Q/\mu_2) + b_2 J_2(Q/\mu_2)\>,
\end{eqnarray}
\end{subequations}
and to select $a_{1,2}$ and $b_{1,2}$ 
such 
that the asymptotic behaviour at large $Q^2$ is preserved. For small
$Q^2$ values this means
\begin{equation}
\label{2.13}
Z_1 \propto \frac{1}{Q}\>,\qquad Z_2 \propto \frac{1}{Q^2}
\end{equation}
and, consequently,
\begin{equation}
\label{2.14}
Z^{-1} \propto \frac{1}{Q^4}\>,\qquad m \propto Q^2\>.
\end{equation}
Instead of the behaviour of $Z^{-1}$ and $m$ corresponding to Fig.~\ref{Fig5}
we now have a behaviour 
as 
shown 
in Fig.~\ref{Fig6}.
\begin{figure}[ht]
\begin{center}
\input{fig6.pstex_t}    
\end{center}
\caption{}
\label{Fig6}
\end{figure}

\noindent
This behaviour corresponds to the confined solution for the quark
Green's function. In this solution the condensate which is created at
momenta of the order of $\lambda$ exists only in a region of $Q^2$
values between $Q_1^2$ and $Q_2^2$ ($Q_1^2,Q_2^2 \sim f^2$) and
disappears at smaller $Q^2$ values due to its decay into $\pi$-mesons.
Because of the decrease of $m(q^2)$ at large and small $Q^2$, in this
solution the pion contribution to the right-hand side of the equation
(\ref{2.4}), $\Delta_{\pi}$, is localised between $Q_1^2$ and $Q_2^2$
(we shall see this in detail later). This localisation of the pion
contribution enables us to analyse not only the behaviour in the
Euclidean region $q^2<0$ but also the analytic properties of the
solution and its behaviour at positive $q^2$ values. 

If the pion is localised, $G^{-1}(q)$ satisfies the old equation
without $\Delta_{\pi}$ at $Q^2 \ll Q_1^2$ and $Q^2 \gg Q_2^2$. This
means that $G^{-1}(q)$ can be written in the form
\begin{equation}
\label{2.15}
G^{-1}(q) \>=\> C_1(q)G_1^{-1}(q) + C_2(q)G_2^{-1}(q)\>,
\end{equation}
where $G_{1,2}^{-1}(q)$ are two different solutions of the equation
(\ref{2.4}) for $\Delta_{\pi}=0$ and $C_{1,2}(q)$ are slowly varying
functions,
\begin{eqnarray*} 
C_1(q) &\rightarrow& \mbox{const} \quad\mbox{at}\quad Q^2 \gg Q_1^2
\quad\mbox{and}\quad C_1(q) \rightarrow 0
\quad\mbox{at}\quad Q^2\ll Q_2^2\>, \\
C_2(q) &\rightarrow& 0 \quad\mbox{at}\quad Q^2 \gg Q_1^2
\quad\mbox{and}\quad C_2(q) \rightarrow \mbox{const} \quad\mbox{at}\quad
Q^2\ll Q_2^2\>.
\end{eqnarray*}
$G^{-1}(q)$ has to satisfy the conditions at large $Q^2$ which we
already imposed on our solution. For this, we are bound to choose
$G_1^{-1}(q)$ to be the solution corresponding to symmetry breaking
with the "mass" $m_c$,
which was discussed in [6], 
and $G_2^{-1}(q)$ has to be the solution describing a singular
behaviour of the type (\ref{2.13}), (\ref{2.14}) at small $Q^2$. We
have to have in mind also that $G^{-1}(q)$ (\ref{2.15}) must not have
singularities in the complex plane.  $G_1^{-1}(q)$ has no
singularities there but the standard cut on the real axis
from $q^2=m_1^2$ to $q^2\rightarrow \infty$. 
We choose $G_2^{-1}$ to have also no singularities in the complex plane.
After that we shall see what will be the singularities of $G^{-1}(q)$.

Making use of the fact that equation (\ref{2.4}) with $\Delta_{\pi}=0$
is scale invariant 
for
$\alpha=\mbox{const}$, we can always write
\begin{equation}
\label{2.16}
G_2^{-1}\left(\hat{q},g\left(q\right)\right) =
 \frac{m_2^4}{q^2}\tilde{G}^{-1}\left(\frac{m_2^2}{\hat{q}},g\left(
\frac{m_2^2}{q}\right)\right) ,
\end{equation}
where $\tilde{G}^{-1}(q)$ satisfies the equation
\begin{equation}
\label{2.17}
\partial^2\tilde{G}^{-1}(q) = g\left(\frac{m_2^2}{q}\right)
\partial_{\mu}\tilde{G}^{-1}(q)\tilde{G}(q)\partial_{\mu}
\tilde{G}^{-1}(q) \>.
\end{equation}
This is the same equation which we have discussed in [6]; the behaviour
of $\tilde{g}(q)\equiv g(m_2^2/q)$ 
is $\tilde{g}\rightarrow 0$ at $q\rightarrow 0$ and
$\tilde{g}(q)\rightarrow g_0$ at $q \rightarrow \infty$, opposite to
that shown in Fig.~\ref{Fig4}. 
It was shown in [3] and [6] that
independently of the behaviour of $g$ we can always find a solution
with a cut along the positive $q^2$ axis from $q^2 = m_2^2$ to
$q^2 \rightarrow \infty$ if we choose the solution not to have
singularities at $q=0$ and if we fix $\tilde{G}^{-1}(q)$ by the
condition $\tilde{G}^{-1}(q)|_{q=0} = m'_2$ ($m_2$ and $m'_2$ are in a
simple relation).

Suppose that $\tilde{G}^{-1}(q)$ is chosen in such a way. Then
$G_2^{-1}(q)$ defined by (\ref{2.16}) will have a cut from $q^2=0$
to $q^2=m_2^2$ and a singular behaviour at $q^2\rightarrow 0$.
As a result, $G^{-1}(q)$ has two cuts in the $q^2$ plane (Fig.~\ref{Fig7}), 
\begin{figure}[ht]
\begin{center}
\input{fig7.pstex_t}    
\end{center}
\caption{}
\label{Fig7}
\end{figure}

\noindent
where we take $m_2>m_1$ since, as we will see in the next section,
this is the only possibility to avoid singularities in the complex
plane; due to reasons to be discussed, $i\varepsilon$ is positioned
in the way shown in Fig.~\ref{Fig7}.

For the sake of simplicity, let us consider the properties of $G^{-1}(q)$
in the $q_0$-plane at the value $\vec{q}=0$ of the space component of $q$.
We will write $G^{-1}(q)$, as in [6], in the form
\begin{equation}
\label{2.18}
G^{-1}(q_0) = G_+^{-1}(q_0)\frac{1+\gamma_0}{2} + G_-^{-1}(q_0)
\frac{1-\gamma_0}{2}\> .
\end{equation}
For example, $G_+^{-1}(q_0)$ has two normal cuts (Fig.~\ref{Fig8}).
\begin{figure}[ht]
 \begin{center}
\input{fig8.pstex_t}    
\end{center}
\caption{}
\label{Fig8}
\end{figure}

\noindent
The singularities 
of $G_{1+}^{-1}(q_0)$
at $q_0=m_1$ and $q_0=-m_1$ are different. It is
clear from 
eq.\ (103) in [6] 
that $G_{1+}^{-1}(q_0=m_1)=0$, while $G_{1-}^{-1}(q_0=m_1)$,
for $g<\frac{1}{2}$,  
is different from zero:  
$G_{1-}^{-1}(q_0=m_1)
= G_{1+}^{-1}(q_0=-m_1)=\mbox{const}$. 

The function $\tilde{G}_{+}^{-1}(q_0)$ also has the analytic
structure of Fig.~\ref{Fig8} with $m_1\to m_2$.
In the course of the 
reflection
$q_0 \rightarrow -{m_2^2}/{q'_0}$
the points on the line $q_0=iQ$ transform into $q'_0 = i{m_2^2}/{Q}$; 
the points $q_0=\pm m_2\mp i\varepsilon$ 
transform into $q'_0= \mp m_2\mp i\varepsilon$,
respectively.
As it follows from all this, $G_+^{-1}$ has singularities corresponding
to Fig.~\ref{Fig10}.
\begin{figure}[ht]
\begin{center}
\input{fig9.pstex_t}    
\end{center}
\caption{}
\label{Fig10}
\end{figure}

These analytic properties are a clear manifestation of the fact that
the Dirac see is destroyed; $G_+^{-1}$ has singularities corresponding
to both positive and negative energies. The parameter $m_2$ has the
meaning of the Fermi energy. In order to find zeros of $G_+^{-1}$
which have to be in the lower $q_0$ half plane and correspond to unstable
quarks with positive and negative energy, we have to know the signs
of $\Im G_{1+}^{-1}$, $\Im G_{2+}^{-1}$ and the signs of $C_1$, $C_2$
in (\ref{2.15}). The solutions $G_{1+}^{-1}(q_0)$ and
$\tilde{G}_+^{-1}(q_0)$ satisfy the normal unitarity condition. Their
imaginary parts at positive $q_0$ have to be negative. This leads to
signs of the imaginary parts as shown in Fig.~\ref{Fig10} (with arrows 
pointing at the positive side of the cuts). 
Having this in mind,
we can write $G_1^{-1}(\hat{q})$ near to its zero (i.e.\ at $q_0$
close to $m_1$) in the form
\begin{equation}
\label{2.19}
G_{1+}^{-1} = (m_1-i\delta-q_0)^{\frac{1}{\beta}}\quad\mbox{,}\quad
\frac{1}{2}<\beta<1\>.
\end{equation}
Since $G_{2+}^{-1}$ 
remains finite
at $q_0\rightarrow m_1$, 
it can be given in the form
\begin{equation}
\label{2.20}
G_{2+}^{-1}(q_0) = \rho e^{-i\phi} \quad\mbox{,}\quad 0<\phi<\pi\> .
\end{equation}
If so,
\begin{equation}
\label{2.21}
G_+^{-1}(q_0) = C_1(m_1-i\delta-q_0)^{\frac{1}{\beta}} + C_2\rho e^{-i\phi}.
\end{equation}
The zeros of $G_+^{-1}(q_0)$ are 
then
defined by the equation
\begin{equation}
\label{2.22}
q_0^* = m_1-i\delta + \left(\frac{\rho C_2}{C_1}\right)^{\beta}
 e^{-i[\beta\phi+(1-\beta)\pi]}\, .
\end{equation}
If $\frac{C_2}{C_1} > 0$, we have  $\Im q_0^* < 0$. In this case the
singularity appears in the lower half plane and describes the unstable
positive energy quark. 
Repeating the calculation for $q_0\rightarrow
- m_2$, we obtain a singularity corresponding to the unstable hole.
Considering $G_-^{-1}(q_0)$ instead of $G_+^{-1}(q_0)$, we find the same
singularities for an anti-quark and an anti-hole.

If the condition $\frac{C_2}{C_1} > 0$ is not satisfied the singularities
can move to the upper half plane. This does not destroy the theory
because they will be on the unphysical sheet. However, in this case
I do not have any simple interpretations for these singularities.

The analytic properties of $G_+^{-1}(q)$ presented in Fig.~\ref{Fig10} imply
that in the limit $\varepsilon \rightarrow 0$ the Feynman Green's
function defined for real $q_0$ values becomes a non-analytic
function. Two analytic functions which have no singularities
in the upper and lower complex half planes can be defined and called
the retarded and the advanced Green's function. At finite $i\varepsilon$
there exists one analytic function with four cuts.
This can be easily seen if we consider $G_+^{-1}(q)$ as a 
function of $q_0$ at a fixed value of the space component $\vec{q}$ of 
the 4-vector $q_{\mu}$. 
In this case we will have Fig.~\ref{Fig11} instead of Fig.~\ref{Fig10}. 
\begin{figure}[ht]
 \begin{center}
\input{fig10.pstex_t}    
\end{center}
\caption{}  
\label{Fig11}
\end{figure}

\noindent
When negative energies are involved, it is natural to expect that
for the exact solution all four cuts have discontinuities different
from zero in the intervals from 
$q_0=|\vec{q}\,|$ to $q_0 \rightarrow \infty$ 
and from $q_0 \rightarrow - \infty$ to $q_0= -|\vec{q}\,|$.

If in the usual, non-confined case we know the Green's function
(calculating it by using Feynman diagrams in the Euclidean space),
we can continue it into positive $q^2$ values and find that the
discontinuity at $q^2> m^2 $ satisfies the unitarity condition. This
means that the retarded Green's function coincides with the Feynman
Green's function.

The confined case is more complicated. The knowledge of the retarded
Green's function in the limit ($i\varepsilon \rightarrow 0$) is not
sufficient for finding the Feynman Green's function. In order to
obtain the Feynman Green's function, we need the equation for the
discontinuities on the new cuts; it will be an equation for the
density matrix of quarks in the vacuum.

In the next section we will find the solution of the equation
(\ref{2.1}) in the limit $i\varepsilon \rightarrow 0$ (retarded
Green's function) without poles and with the properties we have
discussed. In section \ref{IV} we shall obtain the equations for
the discontinuities on the new and old cuts.

\section{Solution for the retarded Green's function \\
  of confined massless quarks}\label{III}

In this section I follow the pattern formulated in the previous
section. I will find the solution of the equation (\ref{2.1}) in
Euclidean space and then, to be sure that it is stable, continue
it into the complex plane. To find the confined solution of the
equation (\ref{2.1}) for the Green's function of light quarks, we will
introduce the same representation for $G^{-1}$ as in [6]:
\begin{equation}
\label{3.1}
G^{-1}(q) = \left(\frac{u}{q}\right)^{\frac{1}{\beta}}e^{-\hat{n}
\frac{\phi}{2}}
\end{equation}
where $\beta=1-g$ and $\hat{n}=\hat{q}/q$. 
Instead of the equations (84), (85) in [6], we will have
\begin{subequations}
\label{3.2}
\begin{eqnarray}
\label{3.2.old}
& & \ddot{\phi} + \frac{2\dot{u}}{u}\dot{\phi} - 3\sinh\phi = 0 \>, \\
\label{3.3}
& & \ddot{u} - u + \left[\,\beta^2\left( 3\sinh^2\frac{\phi}{2} 
+ \frac{\dot{\phi}^2}{4}\right) - \frac{q^2\beta\cosh^2\frac{\phi}{2}} {f^2}
\,\right] u = \frac{\dot{\beta}}{\beta}(\dot{u}-u)\>,
\end{eqnarray}
\end{subequations}
where $\dot{f}= \frac{\partial f}{\partial\xi}$ and $\xi = \ln q$.
As we see, the pion contribution to the equation (\ref{2.1}) influences
only the equation for $u(\xi)$ and it depends explicitly on $q^2$.
Because of this, for the equations (\ref{3.2})
the energy is not conserved even if $\beta$ is constant.  In the
Euclidean region, $q^2=-Q^2<0$, we can write $\phi=i\psi$ and the
equations take the form
\begin{subequations}
\label{3.4.all}
\begin{eqnarray}
\label{3.4}
&& \ddot{\psi} + 2p\dot{\psi} - 3\sin\psi = 0 \>, \\
\label{3.5}
&& \ddot{u} - V(\xi)u = 0 \>,
\end{eqnarray}
\end{subequations}
where
\begin{equation}
\label{3.6}
V(\xi) = 1+\beta^2 \left(3\sin^2\frac{\psi}{2}+\frac{\dot{\psi}^2}{4}
\right) - \frac{Q^2}{f^2} \beta\cos^2\frac{\psi}{2} 
+
 \frac{\dot{\beta}} {\beta}(p-1)
\end{equation}
and $ p = {\dot{u}}/{u}$.
As before, the equation (\ref{3.4}) corresponds to particle propagation
in a periodic potential with damping when $\xi$ is increasing and with
acceleration when $\xi$ is decreasing.
\begin{figure}[ht]
\begin{center}
\input{fig11.pstex_t}    
\end{center}
\caption{}
\label{Fig12}
\end{figure}

\noindent
In [6] we considered the trajectories of the type I and II (dashed
curves); now we will concentrate mainly on the trajectory $C$ (solid
curve) which has the structure we discussed in the previous section.
Any given trajectory defines the potential $V(\xi)$ (if we neglect
$\frac{\dot{\beta}}{\beta}(\dot{u}-u)$) in the equation (\ref{3.5})
which is the Schr\"odinger equation at zero energy. The solution of
this equation, defining the damping 
$p$ 
in (\ref{3.4}), has to be chosen
self-consistently.

The structure of the equations (\ref{3.4.all}) with pion contribution
differs essentially from the structure of the equations without pions
even at large $Q^2$ in the sense that they have no solutions
corresponding to massive quarks. The massless solutions of these
equations are not very different with or without $\pi$-mesons.

The solutions of (\ref{3.4.all}) satisfying the condition
of asymptotic freedom are, at large $Q^2$,
\begin{subequations}
\label{3.8.all}
\begin{eqnarray}
\label{3.8}
\frac{1}{2}(\psi-\pi) &=& \frac{\nu^3}{Q^2}\left(\frac{\alpha}{\alpha_0}
\right)^{-3\gamma} , \\
\label{3.9}
u &=& Q^2 u_0 \left\{\left(\frac{\alpha}{\alpha_0}\right)^{\gamma} +
\frac{\nu_1^4}{Q^4}\left(\frac{\alpha}{\alpha_0}\right)^{-\gamma} -
\frac{\nu^6}{8f^2Q^4}\frac{\pi}{2}\left(\frac{\alpha}{\alpha_0}\right)
^{-2\gamma}\right\} , 
\end{eqnarray}
\end{subequations}
where $\gamma = \frac{4}{b}$ is the invariant anomalous dimension;
$b=\frac{11}{3}N_c-\frac{2}{3}n_f$.
The renormalisation of the Green's function which we discussed in the
previous section is
\begin{equation}
\label{3.10}
Z^{-1}(Q) = u_0\left(\frac{\alpha}{\alpha_0}\right)
^{\gamma/\beta}
\quad\mbox{at}\quad Q\rightarrow \infty
\end{equation}
in Feynman gauge (which we are using here).

It is not clear at all whether it makes sense to take these anomalous
dimensions seriously. For the sake of simplicity, we shall consider
(\ref{3.8.all}) at $\gamma=0$. Important is, that the
solutions of (\ref{3.4.all}) still contain two parameters.

For small $Q^2$ values the non-confined solutions can be written in
the simple form
\begin{subequations}
\label{3.11}
\begin{eqnarray}
\label{3.11.1}
\psi & = & \frac{Q}{m_c}\>, \\
\label{3.11.2}
u & = & Q\left\{1+\frac{1}{8}\beta(0)\left(\beta(0)\frac{Q^2}{m_c^2} -
\frac{Q^2}{f^2}\right)\right\}.
\end{eqnarray}
\end{subequations}
The confined solution has a more complicated structure. For $Q^2
\rightarrow 0$, $\psi \rightarrow \pi$ we have in this case
\begin{subequations}
\label{3.12.all}
\begin{eqnarray}
\frac{\psi-\pi}{2} &\simeq& \left(\frac{Q}{m_2}\right)^p C\cos\left(\sqrt{2-
3\beta^2}\ln\frac{Q}{Q_0}\right) , \\
\label{3.13}
u&\simeq&\left(\frac{m_2}{Q}\right)^p
u_0\left\{1-\left(\frac{Q}{m_2}\right)^{2p}  
\frac{\beta^2 C^2}{p}\ln\frac{Q}{Q_1}\right. \nonumber\\
 & &  \hspace{2cm}+ \left.\left(\frac{Q}{m_2}\right)^{2p}
 \frac{\beta^2 C^2}{\sqrt{2-3\beta^2}}\cos\big(2\sqrt{2-3\beta^2}
 \ln\frac{Q}{Q_0} + \delta\big)\right\},
\end{eqnarray}
\end{subequations}
with $ p=\sqrt{1+3\beta^2}$, $\beta=\beta(0)$. \\
The solutions contain three essential parameters $C$, $Q_0$ and $Q_1$
which define $\tilde{\psi}=\psi-\pi$, $\dot{\tilde{\psi}}$ and $\dot{u}$
at $Q=m_2$; $C={\tilde{\psi}_0}/{2}$,
\begin{equation}
\label{3.14}
\frac{\dot{\tilde{\psi}}_0}{2} = p-\sqrt{2-3\beta^2} \tan \left(
\sqrt{2-3\beta^2} \ln \frac{m_2}{Q_1} \right) .
\end{equation}
Let us now consider the structure of the solution in the intermediate
region. We will start with the qualitative discussion of the
non-confined solution of the type I. In this case the potential
$V_0(\xi)$ in (\ref{3.5}) without pion contribution behaves as shown
in Fig.~\ref{Fig13} (solid line) and the solution $u_0(\xi)$ corresponds to the
dashed curve in the same figure, where $\lambda$ is the QCD scale at
which $\beta^2=2/3$ and $m_c$ is the quark mass.
\begin{figure}[ht]
\begin{center}
\input{fig12.pstex_t}    
\end{center}
\caption{}
\label{Fig13}
\end{figure}

The pion-induced potential 
\begin{equation}
V_{\pi}(\xi) = - \frac{Q^2}{f^2}\beta \cos^2\frac{\psi}{2}
\end{equation}
decreases at large and small $Q^2$ values:
\[ 
V_{\pi}=-\frac{Q^2}{f^2}\beta(0) \quad\mbox{at}\quad Q^2< m_c^2 \>,
\qquad V_{\pi}=-\frac{\nu^6}{f^2 Q^4}\quad\mbox{at}\quad
Q^2>\lambda^2\> .
\]
If $ f \sim m \sim \nu < \lambda$, as it is natural to expect from the
expression 
for $ V_{\pi}$,
the total potential $V$ becomes essentially
different from $V_0(\xi)$, especially for small $\beta(0)$ values. 
It will correspond roughly to the second solid curve in
Fig.~\ref{Fig13}. 
The solution $u(\xi)$ of the equation (\ref{3.5}) also changes
substantially. 
Its exact form depends, of course, on the values of $f$, $\nu$ and on the
behaviour of $\beta(0)$. At sufficiently small $\beta(0)$ it can even
have a shape $u(\xi)$ corresponding to the second dashed line in
Fig.~\ref{Fig13}. 
From the point of view of the structure of the solution in
Euclidean space, these changes do not matter too much (at least at the
first sight). 
But, as we will see, from the point of view of the analytic
continuation the situation 
is 
different.

Let us consider now the behaviour of the potential $V(\xi)$ and the
solution $u(\xi)$ corresponding to the "confined" trajectory $C$. In
this case $V_0(\xi)$ and $u_0(\xi)$ behave as shown in Fig.~\ref{Fig14}. 
\begin{figure}[ht]
\begin{center}
\input{fig13.pstex_t}    
\end{center}
\caption{} 
\label{Fig14}
\end{figure}

\noindent
The pion potential $V_{\pi}(\xi)$ in this case is 
\begin{equation}
\label{3.15}
V_{\pi}(\xi) = \left\{ \begin{array}{cl}
 -\frac{\nu^6}{f^2 Q^4} & \quad\mbox{at}\quad Q>\lambda \\
-\beta\,\frac{Q^2}{f^2}\left(\frac{Q}{m_2}\right)^{2\beta}
\frac{1}{2}
\left[\, 1+\cos \left(2\sqrt{2-3\beta^2}\ln\frac{Q}{Q_0}\right)\right] &
\quad\mbox{at}\quad Q<m_2\>.
\end{array} \right.
\end{equation}
This potential is strongly localised near the value $Q\!=\!Q_0$ where the
trajectory $C$ reaches its highest point (point $\psi_0$ in Fig.~\ref{Fig12}). 
If this highest point is in the region close to zero, 
the minimum of $V_{\pi}(\xi)$ equals
$-{Q_0^2}\beta(Q_0)/f^2$. As a result, the potential $V(\xi)$
has a behaviour shown in 
Fig.~\ref{Fig14}. 
The solution $u(\xi)$ looks qualitatively the same. 
At some point its derivative $\dot{u}(\xi)$ changes
sign, which leads to the transition from acceleration to damping in the
equation (\ref{3.4}). 
As a result, the trajectory returns to the minimum of the well. 
Obviously, there is always a solution for
$u(\xi)$ with such a behaviour. Because of this, a solution of the
type $C$ for $\psi(\xi)$ also exists and depends on two parameters $\nu_1$
and $Q_0$. The problem with this solution is not its existence in
Euclidean space but its behaviour at complex $q^2$.

In order to understand the problem of the analytic continuation, we
have to remember that the singularity of $G^{-1}$ corresponds to the
point where
\[ u(q) = 0 \]
and find out how the position of this zero depends on the structure of
$u(\xi)$ at negative $q^2$.

Let us start with the simple quasi-classical solution for $u$:
\begin{equation}
\label{3.16}
u = u_0 \cosh \zeta(\xi)\quad\mbox{,}\quad \zeta(\xi) = \int_{\xi_0}
^{\xi} \sqrt{V(\xi')}d\xi'\>,
\end{equation}
where $\xi_0$ is chosen so that $\dot{u}(\xi_0)=0$,
and consider the analytic continuation of (\ref{3.16}) along the
circle in the complex plane of 
$q=Q_0e^{i(\frac\pi2-\chi)}$, $\chi\ge0$; $\xi_0=\ln Q_0$ 
(see Fig.~\ref{Fig15}).
\begin{figure}[ht]
  \begin{center}
\input{fig14.pstex_t}    
\end{center}
\caption{}
\label{Fig15}
\end{figure}

\noindent
In this continuation we have
\begin{eqnarray}
\label{3.17}
\zeta(\xi) & = & -i \int_0^{\chi}
\sqrt{V\left(\xi_0+\frac{i\pi}{2}-i\chi'\right)}\>d\chi' 
= \zeta_1-i\zeta_2\>,\nonumber\\
u & = & \cosh\zeta_1\cos\zeta_2 - i\sinh\zeta_1\sin\zeta_2 \>;
\end{eqnarray}
$u$ will have a zero at the point $\zeta_2=\frac{\pi}{2}$, $\zeta_1=0$.
The condition $\zeta_1=0$ means that the singularity appears near the
circle where $\dot{u}(\xi_0) = 0$. The condition $\zeta_2=\frac{\pi}{2}$
gives
\begin{equation}
\label{3.18}
\zeta_2 = 
\chi\sqrt{V(\xi_0)}\>, \qquad 
\chi=\frac{\pi}{2\sqrt{V(\xi_0)}}\> .
\end{equation}
Hence, we will have a singularity in the {\em upper}\/ half-plane if
$V(\xi_0)>1$, and in the {\em lower}\/ half-plane, on the unphysical sheet,
if $V(\xi_0)<1$.

From (\ref{3.6}) it is obvious that if the pion potential is not
present, 
then 
$V(\xi)>1$ and there will be a singularity on the physical
sheet if the solution has a minimum (the acceleration transforms into
damping). But with the pion potential included, $V(\xi_0)$ can be less
than unity, and the singularity will be on the unphysical sheet if we
choose a solution with a minimum close to the minimum of the potential.
At the same time, considering the behaviour of the solution for the
non-confined trajectory (Fig.~\ref{Fig13}), 
we see the opposite situation: without
the pion potential $u_0(\xi)$ has no zeros and thus there is no
reason for singularities. However, if in the presence of the pion
potential at a small $\beta(0)$ the solution $u(\xi)$ shown in Fig.~\ref{Fig13}
exists and has a minimum in the region where $V(\xi)>1$, we will
definitely have a singularity in the upper half plane. It is important
to stress that for the non-confined trajectory the position of the
minimum is defined by the form of the potential and by the condition
according to which the solution has to approach zero in the limit
$Q\rightarrow 0$. This means that at least for 
$g(0)$ 
close to unity the non-confined solution will be unstable.

We shall use this qualitative considerations as a hint how to obtain
constructively the stable solution for the confined trajectory. Let us
consider the trajectory $C$ for which $\psi_0$, defined as the point
where $\dot{u}(\xi_0)=0$, is very small, $\psi\ll 1$, and so is
$\psi_1$ where $\dot{\psi}=0$.  In this case the potential $V(\xi)$ in
(\ref{3.5}) will be
determined 
mainly by the pion potential. In addition, it follows
from (\ref{3.4}) that $\psi(\xi)$ will be a slowly changing function
of $\xi$ near $\xi_0$ where $\dot{\psi}=0$ and $\dot{u}=0$. 
In this region the equation (\ref{3.5}) can be written as
\begin{equation}
\label{3.19}
\ddot{u} - \nu^2 u + x^2\mu^2 u = 0
\end{equation}
where 
$\nu^2 = 1+\beta^2 \left(3\sin^2\frac{\psi}{2}+ \frac{\dot{\psi}^2}{4}\right)$,
$\mu^2=\beta^{-1}\cos^2\frac{\psi}{2}$ 
and $x=\frac{Q}{f}\beta$ with slowly varying
$\nu$ and $\mu$, $\nu^2\approx 1$ and $\mu^2\approx \beta^{-1}$.

For constant $\mu$ and $\nu$ values, (\ref{3.19}) is the Bessel
equation in $x\simeq \ln\xi$.
Its solution can be given in the form
\begin{equation}
\label{3.20}
u = c_1 Y_{\nu}(\mu x) + c_2 J_{\nu}(\mu x) .
\end{equation}
We choose the coefficients $c_1$, $c_2$ in such a way that $\dot{u}
(\xi_0)=0$ at a point 
$x_0={Q_0}{f}/\beta(x_0)$ inside the region
between $Q_1$ and $Q_2$ where $\nu$ 
is
close to unity.

Outside this region $\dot{u}(\xi)\neq 0$. Indeed, 
in the $x\to0$ limit 
only the singular Bessel function is important, 
\begin{equation}
\label{3.21}
u = c_1 Y_{\nu}(\mu x) \sim (\mu x)^{-\nu}\>.
\end{equation}
For large $x$, as it can be seen from (\ref{3.15}), 
$u$ also has a power behaviour (\ref{3.21}) 
(only with different $\mu(Q)$ and $\nu(Q)$).

Thus we conclude that in the dangerous
region where $\dot{u}(\xi)$ can be zero we have an explicit expression
for
\begin{equation}
\label{3.22}
u(\xi) = c_1 Y_1(x) + 
c_2 J_1(x) \>,
\end{equation}
which allows us to carry out the analytic continuation along the
strip shown in Fig.~\ref{Fig15} and to see where the zeros of $u(\xi)$ are.

The most interesting case is when the singularity appears near the
real axis in the $q$ plane: $ 
\chi
=\frac{\pi}{2}+\delta $,
\begin{equation}
\begin{split}
\label{3.23}
J_1(e^{-i\frac{\pi}{2}}x') & =  -iI_1(x') \>,\\
Y_1(e^{-i\frac{\pi}{2}}x') & =  -\frac{2}{\pi}iK_1(x') - I_1(x')\>.
\end{split}
\end{equation}
The equation for the position of the singularity is
\begin{equation}
\label{3.24}
-i\left[ \frac{2}{\pi}K_1(x') - \gamma I_1(x')\right] - I_1(x')=0\>,
\quad 
\gamma= -\frac{c_2}{c_1} 
= \frac{\dot{Y_1}(x_0)} {\dot{J_1}(x_0)}\>.
\end{equation}
It is clear from this equation that there is no singularity on the
real axis. 
Near the real axis, $x'=xe^{-i\delta}\approx x-i\delta x$, we have
\begin{equation}
\begin{split}
\label{3.25}
\frac{2}{\pi}K_1(x) - \gamma I_1(x) - \dot{I}_1(x)\cdot \delta & =  0 
\>, \\
\left( \frac{2}{\pi}\dot{K}_1(x) - \gamma \dot{I}_1(x)
\right)\cdot\delta  
+ I_1(x) & =  0, \quad \dot{f}=x\partial_x f(x) \>.
\end{split}
\end{equation}
If $x$ is not small, there is no reason to expect a singularity near
the real axis since in this case the potential in (\ref{3.19}) is much
less than unity. 
For small $x$ values we have
\begin{equation}
\label{3.26}
\frac{2}{\pi}K_1(x) \>\simeq\> \frac{2}{\pi x} \left(1+\frac{x^2}{2} \ln
\frac{x}{c}\right); \qquad I_1(x)\>\simeq\> \frac{x}{2}\> .
\end{equation}
The $\gamma$ parameter for $x_0\ll1$ becomes 
\begin{equation}
\label{3.28}
 \gamma \equiv \frac{\dot{Y}_1(x_0)}{\dot{J}_1(x_0)}
  = \frac{4}{\pi x_0^2} \gg 1\>, \quad
(\gamma > 1 \quad\mbox{for}\quad x_0<3)\,.
\end{equation}
Using (\ref{3.26}) it is easy to derive the solution of (\ref{3.25}): 
\begin{equation}
\label{3.29}
\delta \simeq \frac{1}{2\gamma}\>, \qquad x^2 \simeq
 \frac{4}{\pi\gamma} \> \approx\> x_0^2\, .
\end{equation}
Hence, the singularity is on the unphysical sheet ($\delta>0$). It is
positioned on the dangerous circle $x=x_0$ and near the real axis when
$x_0$ is small.

Performing the analytic continuation we supposed that $\psi$ is
changing slowly as the function of $q$. This hypothesis is correct, if
the singularity is far away from the real axis ($\delta>1$). If, however,
$\delta\ll 1$, we have to be careful:
near the singularity the trajectory $\phi(\xi)$ 
changes fast, jumping to a large value $\phi\sim \ln 1/\delta$. 
%
     

%
Fortunately, this jump does not influence
essentially the behaviour of $u(\xi)$ or the position of the singularity.
At first sight this jump is important because near the singularity
we have
\begin{equation}
\label{3.30}
u \>\sim\> \sqrt{\mu^2-q^2} \qquad\mbox{,}\qquad \dot{\phi} \sim \frac{1}
 {\mu^2-q^2}\>,
\end{equation}
and in the equation for $u$ two terms $\ddot{u}$ and 
$u{\dot{\phi}^2}/{4}$ are of the same order. 
Let us, however, consider the equation for
$z=u^2$. We can write this equation in the form
\begin{equation}
\label{3.31}
\ddot{z}-2\left(1-3\beta^2\sinh^2\frac{\phi}{2}+\frac{q^2}{2f^2}\beta\cosh^2
\frac{\phi}{2} \right)z - \frac{1}{2}\left(\frac{\dot{z}^2}{z^2}
- \beta^2\dot{\phi}^2\right)z = 0\,.
\end{equation}
It is easy to show that the last term in this equation has a finite
limit $E$  when $q^2\rightarrow \mu^2$. Hence, the equation can be
rewritten as
\begin{equation}
\label{3.32}
\ddot{z} - 2\left(1-3\beta^2 \sinh^2\frac{\phi}{2} + \frac{q^2}{2f^2}
\beta\cosh^2 \frac{\phi}{2} \right)z + E = 0\,.
\end{equation}
Its solution near the singularity is
\begin{equation}
\label{3.33}
z=(q^2-\mu^2) + c(q^2-\mu^2)^{3-\frac{1}{\beta}} \quad\mbox{;}\quad
e^{\phi} \sim (\mu^2-q^2)^{-1/\beta}
\end{equation}
and the correction to $z=u^2$ from the jump in the trajectory turns
out to be small if $\beta>\frac{1}{2}$.

\section{Analytic properties of the Green's functions \\
  of confined quarks}\label{IV}

In the previous sections we have shown that the equations for
Green's functions are not sufficient to define the theory. It is
easy to show that without confinement the requirement of
usual analytic properties for the Green's functions is
enough for defining the solution for the Green's functions. But even in
this case it is not clear a priori, what kind of bound states have to
be introduced and how these states would influence the equations for
Green's functions. In particular, we have to see how Goldstone type
states which necessarily have to exist in some cases will affect the
equations. This question will be discussed later. Having it in mind, we
first have to learn what kind of analytic properties are necessary
for the Green's functions of quarks and gluons if there is confinement.

Let us start with the usual definition (given by Dyson) for the
Green's function of the quark:
\begin{equation}
\begin{split}
\label{4.1}
G_D(x-x')& =\langle T\Psi(x)\bar{\Psi}(x') \rangle  \\
 & =\vartheta(x_0-x'_0)\langle\Psi(x)\bar{\Psi}(x')\rangle -
 \vartheta(x'_0-x_0)\langle\bar{\Psi}(x')\Psi(x)\rangle .
\end{split}
\end{equation}
The quantity on the right-hand side can be written in the form
\begin{equation}
\label{4.2}
\langle\Psi(x)\bar{\Psi}(x')\rangle = \int\frac{d^4 p}{(2\pi)^4}
 f(p)e^{-ip(x-x')}\>,
\end{equation}
where
\begin{equation}
\label{4.2'}
\overline{f(p)} = \gamma_0 f^+(p) \gamma_0 = f(p)\>,
\end{equation}
\begin{equation}
\label{4.2''}
 f^+(p) = \sum_n \langle 0|\Psi(0)|n\rangle
  \langle n|\bar{\Psi}(0)|0 \rangle\> \delta(p-p_n)\>,
\end{equation}
and
\begin{equation}
\label{4.3}
\langle\bar{\Psi}(x')\Psi(x)\rangle = \int\frac{d^4 p}{(2\pi)^4}
 \tilde{f}(p)e^{-ip(x-x')} \>;
\end{equation}
\begin{equation}
\label{4.3'}
\overline{\tilde{f}}(p) = \tilde{f}(p)\>,
\end{equation}
\begin{equation}
\label{4.3''}
\tilde{f}(p) = \sum_n \langle 0|\bar{\Psi}(0)|n \rangle
  \langle n|\Psi(0)|0 \rangle \>\delta(p+p_n)\>.
\end{equation}
Making use of charge conjugation, we can see that
\begin{equation}
\label{4.4}
\tilde{f}(p) = C^{-1}f(-p)C \>.
\end{equation}
The usual hypothesis which leads to simple analytic properties
for $G(p)$ is
\begin{equation}
\label{4.5}
p_n^2 \geq 0 \>, \qquad p_{n0}> 0 \>,
\end{equation}
for the energy-momentum of the intermediate states $p_n$ in (\ref{4.2})
and (\ref{4.3}). If this were true the quarks would be observable.

If there are no stationary states with quark quantum numbers, the
functions $f(p)$ and $\tilde{f}(p)$ are still related by (\ref{4.4})
and they still satisfy (\ref{4.2'}) and (\ref{4.3'}) but no other
simple properties. It can be said only that the Dyson $T$-product
$G_D$ can be written in momentum space as the sum of two analytic
functions $G_1$, $G_2$:
\begin{subequations}
\label{4.6.all}
\begin{eqnarray}
\label{4.6}
G_D &=& G_1 - G_2\>, \\
\label{4.7}
G_1(p_0,\vec{p}) &=& \frac{1}{\pi}\int_{-\infty}^{\infty}\frac{dp'_0}
{p'_0-p_0-i\varepsilon}\> f(p'_0,\vec{p})\> , \\
\label{4.8}
G_2(p_0,\vec{p}) &=& \frac{1}{\pi}\int_{-\infty}^{\infty}\frac{dp'_0}
{p'_0-p_0+i\varepsilon}\> \tilde{f}(p'_0,\vec{p})
\end{eqnarray}
\end{subequations}
Nevertheless, we will see that the requirements of causality and
unitarity impose severe restrictions on the functions $f(p)$ and
$\tilde{f}(p)$.

Let us regard first the causality. In order to exploit it, consider
the retarded Green's function
\begin{equation}
\label{4.9}
G_R(x-x') = \vartheta(x_0-x_0')\left\{\Psi(x),\overline{\Psi}(x')\right\} \>.
\end{equation}
In momentum space $G_R$ can be written in terms of the same functions
$f(p)$ and $\tilde{f}(p)$
\begin{equation}
\label{4.10}
G_R(p_0,\vec{p}) = G_1(p_0,\vec{p}) + G^*_2(p_0,\vec{p})\>,
\end{equation}
where
\begin{equation}
\label{4.11}
G^*_2(p_0,\vec{p}) = \frac{1}{\pi}\int_{-\infty}^{\infty} \frac{dp'_0}
 {p'_0-p_0-i\varepsilon} \>\tilde{f}(p'_0,\vec{p}) \>.
\end{equation}
The Green's function (\ref{4.9}) has to be zero for $(x-x')^2<0$.
This condition can be formulated in terms of $f(p)$ and $\tilde{f}(p)$
as
\begin{equation}
\label{4.12}
f(p)=-\tilde{f}(p)\quad\mbox{ at }\quad p^2<0 \>.
\end{equation}
The results (\ref{4.6.all}) and (\ref{4.10})-(\ref{4.12}) can
be expressed in terms of one analytic function in the complex plane
with cuts as shown in Fig.~\ref{Fig17}.
\begin{figure}[ht]
\begin{center}
\input{fig15.pstex_t}    
\end{center}
\caption{}
\label{Fig17}
\end{figure}

\noindent
The Dyson Green's function is defined for real $p_0$ values and has
an imaginary part different from zero everywhere. The retarded Green's
function is defined in the upper half plane and, due to the condition
(\ref{4.12}), has no imaginary part between $-|\vec{p}\,|$ and $|\vec{p}\,|$.
This means that $G_R$ can be defined in the complex plane with four
cuts (Fig.~\ref{Fig18}) provided $i\varepsilon$ is finite.
\begin{figure}[ht]
\begin{center}
\input{fig16.pstex_t}    
\end{center}
\caption{}
\label{Fig18}
\end{figure}

\noindent
In order to find the analytic properties corresponding to Fig.~\ref{Fig18},
the cuts denoted by the dashed lines have been moved from the region
$-|\vec{p}\,|<p_0<|\vec{p}\,|$ to the regions $p_0>|\vec{p}\,|$ and
$p_0<-|\vec{p}\,|$. Of course, this procedure defines a new function,
different from $G_D$, on the real axis; it is natural to call it the
Feynman Green's function $G_F$. It is, I believe, possible to show that
this function corresponds to the Fourier component of
\begin{equation}
\label{4.13}
G_F(x,x') = \langle TS\Psi(x)\bar{\Psi}(x')\rangle
\end{equation}
where $S$ is the $S$-matrix in interaction representation; $|\rangle$,
$\langle|$ are free vacuum states. $G_F(x-x')$ is usually calculated
as a series of Feynman diagrams. Generally speaking, (\ref{4.1}) is not
equal to (\ref{4.13}).

Independently of introducing $G_F$, Fig.~\ref{Fig18} 
suggests that we can
define two functions $G_+$, $G_-$. $G_+(p)$ is the contribution of the
usual cuts ($|\vec{p}\,|-i\varepsilon,\infty$), ($-\infty,-|\vec{p}\,|+
 i\varepsilon$)
\begin{equation}
\begin{split}
\label{4.14}
G_+(p) &=  \frac{1}{\pi}\int_{|\vec{p}|-i\varepsilon}^{\infty}
 \frac{dp'_0 \Delta_+(p_0,\vec{p})}{p'_0-p_0} -
\frac{1}{\pi}\int_{-\infty}^{-|\vec{p}|+i\varepsilon}
 \frac{dp'_0 \Delta_+(-p_0,\vec{p})}{p'_0-p_0}  \\
&= \frac{1}{\pi} \int_{0-i\varepsilon}^{\infty} \frac{dp^{'2}}{p^{'2}
 -p^2} \Delta_+(p^{'2})
\end{split}
\end{equation}
and $G_-(p)$ is the contribution of the new cuts
\begin{equation}
\label{4.15}
G_-(p) = \frac{1}{\pi} \int_{0+i\varepsilon}^{\infty} \frac{dp^{'2}}
{p^{'2}-p^2} \Delta_-(p^{'2}) .
\end{equation}
The expressions (\ref{4.14}) and (\ref{4.15}) have to be understood
in the usual way, as equations for two invariant functions. The
discontinuities $\Delta_+$ and $\Delta_-$ are now complex but
$\Delta_+ + \Delta_-$ is real. The unitarity relations which under
the usual (non-confined) circumstances define the discontinuities on
the normal cuts have to define also the new discontinuities. To find
them is equivalent to finding the functions $f(p)$
and $\tilde{f}(p)$ which are the imaginary parts of the Dyson Green's
function and the retarded Green's function, respectively:
\begin{subequations}
\begin{eqnarray}
\label{4.16}
\Im G_D(p) &=& f(p) - \tilde{f}(p)\> , \\
\label{4.17}
\Im G_R(p) &=& f(p) + \tilde{f}(p)\> .
\end{eqnarray}
\end{subequations}

\section{Unitarity relations for quark and gluon \\
Green's functions}
\label{V}

Let us try to calculate $\langle \Psi(x)\bar{\Psi}(x')\rangle$
using the hypothesis that the interaction can be introduced
adiabatically in the theory. This idea is not obvious at all. In
the normal perturbative case it means that starting with a
free theory and introducing the interaction adiabatically, we
gradually reach the vacuum state corresponding to a theory
which includes interaction (curve $a$ in Fig.~\ref{Fig19}).
\begin{figure}[ht]
\begin{center}
\input{fig17.pstex_t}    
\end{center}
\caption{}
\label{Fig19}
\end{figure}

\noindent
But if the spectrum of the theory with interaction differs
drastically from the spectrum of the free theory, this hypothesis will
not work because there has to be level crossing. The curve which has
a chance to reach the physical vacuum (curve $b$ in Fig.~\ref{Fig19}) starts
not at a vacuum state with $g=0$ but at an excited state. In our
case the adiabatic hypothesis is equivalent to the assumption that
there exist some excited states in the free theory from which the
physical vacuum can be reached by introducing the interaction
adiabatically.

In order to calculate $\langle \Psi(x)\bar{\Psi}(x') \rangle$, let us
consider the Dirac equation for the quark field $\Psi(x)$
\begin{equation}
\label{5.1}
(\hat{\partial}-\hat{A})\Psi - m\Psi = 0 ;
\end{equation}
here $\hat{A}$ is the operator of the potential for the colour field.
The operator $\Psi$ can be written in terms of its initial conditions
at $t\rightarrow -\infty$ in the form
\begin{equation}
\begin{split}
\label{5.2}
\Psi(x) & =  \int_{y_0\rightarrow -\infty}d^3 y \left[\,\partial_0 G^0_R(x-y)
 \Psi(y) + G^0_R(x-y)\partial_0 \Psi(y)\,\right]  \\
 & \hspace{4mm}
 +  \int d^4 x' \,G^0_R(x-x')\hat{A}\Psi(x')\>,
\end{split}
\end{equation}
where $G_R$ is the retarded Green's function of the free Dirac
equation. Diagrammatically (\ref{5.2}) can be expressed in the form
\begin{equation}
\label{5.3}
\input{fig18.pstex_t} + \input{fig19.pstex_t} + \input{fig20.pstex_t}
+ \cdots
\end{equation}
Expressing $A$ through its initial conditions we will have, for
example, the diagram
\begin{equation}
\label{5.4}
\input{fig21.pstex_t}
\end{equation}
instead of the second diagram in (\ref{5.3}); the dashed line
corresponds to the solution of the free equation for the gluonic
potential. Doing now the same for $\bar{\Psi}(x')$ with the advanced
Green's function, we obtain for $\Psi(x)\bar{\Psi}(x')$ the
following diagrammatic structure, where $y_0=y'_0$:
\begin{equation}
\label{5.5}
\input{fig22.pstex_t} + \input{fig23.pstex_t} + \input{fig24.pstex_t}+ \cdots
\end{equation}
Let us denote
\begin{equation}
\label{5.6}
\begin{split}
\langle \Psi(x)\bar{\Psi}(x') \rangle & =  \begin{picture}(0,0)%
\epsfig{file=fig25.pstex}%
\end{picture}%
\setlength{\unitlength}{4144sp}%
\begingroup\makeatletter\ifx\SetFigFont\undefined%
\gdef\SetFigFont#1#2#3#4#5{%
  \reset@font\fontsize{#1}{#2pt}%
  \fontfamily{#3}\fontseries{#4}\fontshape{#5}%
  \selectfont}%
\fi\endgroup%
\begin{picture}(1374,114)(1339,-2143)
\end{picture}
 \\
\langle A(x)A(x') \rangle & =  \begin{picture}(0,0)%
\epsfig{file=fig26.pstex}%
\end{picture}%
\setlength{\unitlength}{4144sp}%
\begingroup\makeatletter\ifx\SetFigFont\undefined%
\gdef\SetFigFont#1#2#3#4#5{%
  \reset@font\fontsize{#1}{#2pt}%
  \fontfamily{#3}\fontseries{#4}\fontshape{#5}%
  \selectfont}%
\fi\endgroup%
\begin{picture}(1374,114)(1339,-2368)
\end{picture}
\, .  
\end{split}
\end{equation}
In order to calculate them, we have to take all possible average
values of $AA$ and $\Psi\bar{\Psi}$. 
As a result, for $\langle \Psi(x)\bar{\Psi}(x') \rangle$ we will have the
equation
\begin{equation}
\label{5.7}
\begin{picture}(0,0)%
\epsfig{file=fig27.pstex}%
\end{picture}%
\setlength{\unitlength}{4144sp}%
\begingroup\makeatletter\ifx\SetFigFont\undefined%
\gdef\SetFigFont#1#2#3#4#5{%
  \reset@font\fontsize{#1}{#2pt}%
  \fontfamily{#3}\fontseries{#4}\fontshape{#5}%
  \selectfont}%
\fi\endgroup%
\begin{picture}(924,452)(1564,-2312)
\put(1576,-2266){\makebox(0,0)[lb]{\smash{\SetFigFont{12}{14.4}{\rmdefault}{\mddefault}{\updefault}$x$}}}
\put(2431,-2266){\makebox(0,0)[lb]{\smash{\SetFigFont{12}{14.4}{\rmdefault}{\mddefault}{\updefault}$x'$}}}
\end{picture}
\,= \,\input{fig28.pstex_t}\,
  +\,\input{fig29.pstex_t}\,+ \, \input{fig30.pstex_t}\,+\cdots\>,
\end{equation}
and a similar graphic equation for
$\langle A(x)A(x') \rangle$. 
The functions corresponding to the vertical lines are now exact
retarded Green's functions
\begin{equation}
\label{5.8}
\begin{picture}(0,0)%
\epsfig{file=fig31.pstex}%
\end{picture}%
\setlength{\unitlength}{4144sp}%
\begingroup\makeatletter\ifx\SetFigFont\undefined%
\gdef\SetFigFont#1#2#3#4#5{%
  \reset@font\fontsize{#1}{#2pt}%
  \fontfamily{#3}\fontseries{#4}\fontshape{#5}%
  \selectfont}%
\fi\endgroup%
\begin{picture}(789,452)(1654,-2312)
\put(1981,-2311){\makebox(0,0)[lb]{\smash{\SetFigFont{12}{14.4}{\rmdefault}{\mddefault}{\updefault}$G_R$}}}
\end{picture}
\,= \,\begin{picture}(0,0)%
\epsfig{file=fig32.pstex}%
\end{picture}%
\setlength{\unitlength}{4144sp}%
\begingroup\makeatletter\ifx\SetFigFont\undefined%
\gdef\SetFigFont#1#2#3#4#5{%
  \reset@font\fontsize{#1}{#2pt}%
  \fontfamily{#3}\fontseries{#4}\fontshape{#5}%
  \selectfont}%
\fi\endgroup%
\begin{picture}(789,452)(1654,-2312)
\put(1981,-2311){\makebox(0,0)[lb]{\smash{\SetFigFont{12}{14.4}{\rmdefault}{\mddefault}{\updefault}$G_R^0$}}}
\end{picture}
\,
  + \,\begin{picture}(0,0)%
\epsfig{file=fig33.pstex}%
\end{picture}%
\setlength{\unitlength}{4144sp}%
\begingroup\makeatletter\ifx\SetFigFont\undefined%
\gdef\SetFigFont#1#2#3#4#5{%
  \reset@font\fontsize{#1}{#2pt}%
  \fontfamily{#3}\fontseries{#4}\fontshape{#5}%
  \selectfont}%
\fi\endgroup%
\begin{picture}(924,463)(1564,-2312)
\put(1981,-2266){\makebox(0,0)[lb]{\smash{\SetFigFont{12}{14.4}{\rmdefault}{\mddefault}{\updefault}$G_R^0$}}}
\end{picture}
\, + \,\begin{picture}(0,0)%
\epsfig{file=fig34.pstex}%
\end{picture}%
\setlength{\unitlength}{4144sp}%
\begingroup\makeatletter\ifx\SetFigFont\undefined%
\gdef\SetFigFont#1#2#3#4#5{%
  \reset@font\fontsize{#1}{#2pt}%
  \fontfamily{#3}\fontseries{#4}\fontshape{#5}%
  \selectfont}%
\fi\endgroup%
\begin{picture}(924,452)(1564,-2312)
\put(2296,-2311){\makebox(0,0)[lb]{\smash{\SetFigFont{12}{14.4}{\rmdefault}{\mddefault}{\updefault}$D_R^0$}}}
\end{picture}
\,+\,
\cdots \hspace{0.5cm}
\end{equation}
Let us see, what happens to (\ref{5.7}) when $y_0 =t\to
-\infty$.
In a theory without confinement the answer is very simple.
If the retarded Green's function has a pole at $p^2=m^2$, the
diagram
\begin{equation}
\label{5.9}
\input{fig35.pstex_t} 
\end{equation}
has a finite limit when $y_0\to - \infty$ and equals
\begin{equation}
\label{5.10}
\int dq_0 \, e^{-iq_0\left(\frac{x_0+x'_0}{2}-y_0\right)}
\int e^{ik(x-x')}\frac{d^4 k}{(2\pi)^2}\>\delta(k^2-m^2)
\>\delta((q-k)^2-m^2)\> f(k) \>,
\end{equation}
where $f(k)$ corresponds to \begin{picture}(0,0)%
\epsfig{file=fig36.pstex}%
\end{picture}%
\setlength{\unitlength}{4144sp}%
\begingroup\makeatletter\ifx\SetFigFont\undefined%
\gdef\SetFigFont#1#2#3#4#5{%
  \reset@font\fontsize{#1}{#2pt}%
  \fontfamily{#3}\fontseries{#4}\fontshape{#5}%
  \selectfont}%
\fi\endgroup%
\begin{picture}(744,114)(1609,-2143)
\end{picture}
\,.
The product of the $\delta$-functions gives
\begin{equation}
\label{5.11}
\delta((q-k)^2-m^2)\,\delta(k^2-m^2) 
  =  \delta(k^2-m^2)\,\delta(q_0^2-2k_0 q_0) \>\>\rightarrow \>\>
\frac{\delta(k^2-m^2)}{2k_0}\delta(q_0)\> ,
\end{equation}
and therefore (\ref{5.10}) does not depend on $y_0$. For a free Green's
function
\begin{equation}
\label{5.12}
\begin{picture}(0,0)%
\epsfig{file=fig37.pstex}%
\end{picture}%
\setlength{\unitlength}{4144sp}%
\begingroup\makeatletter\ifx\SetFigFont\undefined%
\gdef\SetFigFont#1#2#3#4#5{%
  \reset@font\fontsize{#1}{#2pt}%
  \fontfamily{#3}\fontseries{#4}\fontshape{#5}%
  \selectfont}%
\fi\endgroup%
\begin{picture}(777,1047)(1114,-2818)
\put(1891,-1906){\makebox(0,0)[lb]{\smash{\SetFigFont{12}{14.4}{\rmdefault}{\mddefault}{\updefault}$x'$}}}
\put(1171,-1906){\makebox(0,0)[lb]{\smash{\SetFigFont{12}{14.4}{\rmdefault}{\mddefault}{\updefault}$x$}}}
\end{picture}
 \,=\,\begin{picture}(0,0)%
\epsfig{file=fig38.pstex}%
\end{picture}%
\setlength{\unitlength}{4144sp}%
\begingroup\makeatletter\ifx\SetFigFont\undefined%
\gdef\SetFigFont#1#2#3#4#5{%
  \reset@font\fontsize{#1}{#2pt}%
  \fontfamily{#3}\fontseries{#4}\fontshape{#5}%
  \selectfont}%
\fi\endgroup%
\begin{picture}(1014,114)(1519,-2143)
\end{picture}
 
\end{equation}
and instead of (\ref{5.7}), we have the usual relation for the
imaginary part of the fermion Green's function
\begin{equation}
\label{5.13}
\begin{picture}(0,0)%
\epsfig{file=fig39.pstex}%
\end{picture}%
\setlength{\unitlength}{4144sp}%
\begingroup\makeatletter\ifx\SetFigFont\undefined%
\gdef\SetFigFont#1#2#3#4#5{%
  \reset@font\fontsize{#1}{#2pt}%
  \fontfamily{#3}\fontseries{#4}\fontshape{#5}%
  \selectfont}%
\fi\endgroup%
\begin{picture}(924,493)(1564,-2312)
\put(1981,-1951){\makebox(0,0)[lb]{\smash{\SetFigFont{12}{14.4}{\rmdefault}{\mddefault}{\updefault}$f(p)$}}}
\end{picture}
 \,=\, \begin{picture}(0,0)%
\epsfig{file=fig40.pstex}%
\end{picture}%
\setlength{\unitlength}{4144sp}%
\begingroup\makeatletter\ifx\SetFigFont\undefined%
\gdef\SetFigFont#1#2#3#4#5{%
  \reset@font\fontsize{#1}{#2pt}%
  \fontfamily{#3}\fontseries{#4}\fontshape{#5}%
  \selectfont}%
\fi\endgroup%
\begin{picture}(924,493)(1564,-2312)
\put(1981,-1951){\makebox(0,0)[lb]{\smash{\SetFigFont{12}{14.4}{\rmdefault}{\mddefault}{\updefault}$f_0(p)$}}}
\end{picture}
 \,
  +\, \begin{picture}(0,0)%
\epsfig{file=fig41.pstex}%
\end{picture}%
\setlength{\unitlength}{4144sp}%
\begingroup\makeatletter\ifx\SetFigFont\undefined%
\gdef\SetFigFont#1#2#3#4#5{%
  \reset@font\fontsize{#1}{#2pt}%
  \fontfamily{#3}\fontseries{#4}\fontshape{#5}%
  \selectfont}%
\fi\endgroup%
\begin{picture}(924,511)(1564,-2360)
\put(1981,-2311){\makebox(0,0)[lb]{\smash{\SetFigFont{12}{14.4}{\rmdefault}{\mddefault}{\updefault}$f(p)$}}}
\end{picture}
\,+\, \cdots\hspace{0.4cm}.
\end{equation}
Repeating the same calculation in a theory in which the Green's
functions of quarks and gluons have no poles but soft singularities
we will get $\langle\Psi(x)\bar{\Psi}(x')\rangle=0$ in the
$t\rightarrow -\infty$ limit. This, of course, cannot be the correct
answer --- in the theory with confinement there exist stable hadrons in
general, and those belonging to the pseudoscalar octet in particular.
The problem can be resolved by introducing these pseudoscalar
particles from the very beginning as elementary objects [6]. In this
case we include in the Dirac equation for $\Psi$ not only
the colour field $A$ but also the pseudoscalar field $\varphi$.
This will add to (\ref{5.7}) diagrams of the form
\begin{equation}
\label{5.14}
\input{fig42.pstex_t} \,+\,\input{fig43.pstex_t} 
\end{equation}
where the wavy lines correspond to pseudoscalar particles; for the
sake of simplicity, we shall call them pions. They satisfy the
equality
\begin{equation}
\label{5.15}
\input{fig44.pstex_t}  \,=\, \begin{picture}(0,0)%
\epsfig{file=fig45.pstex}%
\end{picture}%
\setlength{\unitlength}{4144sp}%
\begingroup\makeatletter\ifx\SetFigFont\undefined%
\gdef\SetFigFont#1#2#3#4#5{%
  \reset@font\fontsize{#1}{#2pt}%
  \fontfamily{#3}\fontseries{#4}\fontshape{#5}%
  \selectfont}%
\fi\endgroup%
\begin{picture}(508,362)(855,-1142)
\put(856,-1141){\makebox(0,0)[lb]{\smash{\SetFigFont{12}{14.4}{\rmdefault}{\mddefault}{\updefault}$x$}}}
\put(1306,-1141){\makebox(0,0)[lb]{\smash{\SetFigFont{12}{14.4}{\rmdefault}{\mddefault}{\updefault}$x'$}}}
\end{picture}

\end{equation}
Instead of (\ref{5.13}), we have
\begin{equation}
\label{5.16}
\begin{picture}(0,0)%
\epsfig{file=fig46.pstex}%
\end{picture}%
\setlength{\unitlength}{4144sp}%
\begingroup\makeatletter\ifx\SetFigFont\undefined%
\gdef\SetFigFont#1#2#3#4#5{%
  \reset@font\fontsize{#1}{#2pt}%
  \fontfamily{#3}\fontseries{#4}\fontshape{#5}%
  \selectfont}%
\fi\endgroup%
\begin{picture}(924,493)(1564,-2312)
\put(1936,-1951){\makebox(0,0)[lb]{\smash{\SetFigFont{12}{14.4}{\rmdefault}{\mddefault}{\updefault}$f(p)$}}}
\end{picture}
\,=\,  \begin{picture}(0,0)%
\epsfig{file=fig47.pstex}%
\end{picture}%
\setlength{\unitlength}{4144sp}%
\begingroup\makeatletter\ifx\SetFigFont\undefined%
\gdef\SetFigFont#1#2#3#4#5{%
  \reset@font\fontsize{#1}{#2pt}%
  \fontfamily{#3}\fontseries{#4}\fontshape{#5}%
  \selectfont}%
\fi\endgroup%
\begin{picture}(1104,540)(1474,-2359)
\put(1936,-1951){\makebox(0,0)[lb]{\smash{\SetFigFont{12}{14.4}{\rmdefault}{\mddefault}{\updefault}$f(p')$}}}
\end{picture}
\,
+\, \begin{picture}(0,0)%
\epsfig{file=fig48.pstex}%
\end{picture}%
\setlength{\unitlength}{4144sp}%
\begingroup\makeatletter\ifx\SetFigFont\undefined%
\gdef\SetFigFont#1#2#3#4#5{%
  \reset@font\fontsize{#1}{#2pt}%
  \fontfamily{#3}\fontseries{#4}\fontshape{#5}%
  \selectfont}%
\fi\endgroup%
\begin{picture}(1104,862)(1474,-2542)
\put(1936,-1951){\makebox(0,0)[lb]{\smash{\SetFigFont{12}{14.4}{\rmdefault}{\mddefault}{\updefault}$f(p')$}}}
\end{picture}
 \,+\, \cdots
\end{equation}
and for the gluonic imaginary part, correspondingly,
\begin{equation}
\label{5.17}
\begin{picture}(0,0)%
\epsfig{file=fig49.pstex}%
\end{picture}%
\setlength{\unitlength}{4144sp}%
\begingroup\makeatletter\ifx\SetFigFont\undefined%
\gdef\SetFigFont#1#2#3#4#5{%
  \reset@font\fontsize{#1}{#2pt}%
  \fontfamily{#3}\fontseries{#4}\fontshape{#5}%
  \selectfont}%
\fi\endgroup%
\begin{picture}(924,493)(1564,-2312)
\put(1936,-1951){\makebox(0,0)[lb]{\smash{\SetFigFont{12}{14.4}{\rmdefault}{\mddefault}{\updefault}$f_g(p)$}}}
\end{picture}
\,=\, \begin{picture}(0,0)%
\epsfig{file=fig50.pstex}%
\end{picture}%
\setlength{\unitlength}{4144sp}%
\begingroup\makeatletter\ifx\SetFigFont\undefined%
\gdef\SetFigFont#1#2#3#4#5{%
  \reset@font\fontsize{#1}{#2pt}%
  \fontfamily{#3}\fontseries{#4}\fontshape{#5}%
  \selectfont}%
\fi\endgroup%
\begin{picture}(1104,637)(1474,-2542)
\end{picture}
\,
  +\, \begin{picture}(0,0)%
\epsfig{file=fig51.pstex}%
\end{picture}%
\setlength{\unitlength}{4144sp}%
\begingroup\makeatletter\ifx\SetFigFont\undefined%
\gdef\SetFigFont#1#2#3#4#5{%
  \reset@font\fontsize{#1}{#2pt}%
  \fontfamily{#3}\fontseries{#4}\fontshape{#5}%
  \selectfont}%
\fi\endgroup%
\begin{picture}(1104,907)(1474,-2542)
\end{picture}
\,+\, \cdots \hspace{0.4cm}.
\end{equation}
The equation (\ref{5.16}) differs essentially from (\ref{5.13}). It
contains no driving term (like $f_0$ in (\ref{5.13})); it is a
non-linear but homogeneous equation. The equation (\ref{5.17})
contains driving terms coming from fermions.

Let us consider in detail (\ref{5.16}) in terms of the equation
(\ref{4.6.all}) for $f(p)$. We can write $f(p)$ in the form
\begin{equation}
\label{5.18}
f(p) = \vartheta(p^2)\{f_+(p) + f_-(p)\} + \vartheta(-p^2)f_{Eucl}(p)\>,
\end{equation}
where $f_{\pm}$ are the contributions of positive and negative
frequencies $p_0$. 

Let us first investigate the contributions
of the functions $f_+$ and $f_-$ to the first term in the right-hand
side of (\ref{5.16}).
The function $f_+(p)$ gives the usual contribution to the imaginary
part of the quark Green's function coming from the intermediate
[positive energy]
$\pi$ and quark ($q_+$) states.
The contribution of $f_-(p)$ is very interesting because the signs
of the particle energies in the intermediate states are different
[$\pi$ and a negative energy quark $q_-$].
The expression for this contribution (without the external legs) is
\begin{equation}
\label{5.19}
 \Delta_{q_-,\pi} = g^2\int \frac{d^4 q}{8\pi^2}[\hat{q}a_-(q^2)+
 b_-(q^2)]\vartheta(-q_0)\delta((p-q)^2-\mu^2)\vartheta(p_0-q_0).
\end{equation}
Here $g$ is the coupling constant connecting the $\pi$-meson to
quarks; $a_-$ and $b_-$ are two invariant functions for the
imaginary part of the quark Green's function.

In order to understand the structure of (\ref{5.19}) let us consider
the integrals
\begin{equation}
\label{5.20}
\begin{split}
I_2(p,\kappa,\mu)
 & =  \int \frac{d^4 q}{8\pi^2}\delta(q^2-\kappa^2)\delta((p-q)^2
 -\mu^2)\vartheta(-q_0)\vartheta(p_0-q_0) \\
 & =  \frac{1}{4\pi}\frac{1}{p^2}\sqrt{[p^2-(\kappa+\mu)^2]
 [p^2-(\kappa-\mu)^2]} \\
  &\hspace{4mm} 
 \times  \vartheta(p^2)\vartheta((\kappa-\mu)^2-p^2)
 \{ \vartheta(-p_0)\vartheta(\kappa-\mu) + \vartheta(p_0)\vartheta
 (\mu-\kappa)\}
  \end{split}
\end{equation}
and
\begin{equation}
\begin{split}
\label{5.21}
I_1(p,\kappa,\mu) 
 & = \int \frac{d^4 q}{8\pi^2}\hat{q}\delta(q^2-\kappa^2)
 \delta((p-q)^2-\mu^2)\vartheta(-q_0)\vartheta(p_0-q_0) \\
 & =  \frac{1}{4\pi}\frac{\hat{p}}{p^4}(p^2+\kappa^2-\mu^2)
 \sqrt{[(\kappa-\mu)^2-p^2][(\kappa+\mu)^2-p^2]} \\
 & \hspace{4mm} 
 \times  \vartheta(p^2)\vartheta((\kappa-\mu)^2-p^2)
 \left\{ \vartheta(-p_0)\vartheta(\kappa-\mu) + \vartheta(p_0)\vartheta
 (\mu-\kappa)\right\} .
\end{split}
\end{equation}
Hence,
\begin{equation}
\label{5.22}
\Delta_{q_-,\pi} = g^2\int d\kappa^2 \{a_-(\kappa^2)I_1(p,\kappa,\mu)
 + b_-(\kappa^2)I_2(p,\kappa,\mu)\} .
\end{equation}
The structure of $I_1$ and $I_2$ implies that $f_-$ on the
right-hand side gives contributions to both $f_-$ and $f_+$ on the
left-hand side. The equations for $f_-$ and $f_+$ can be written
in the following way:
\begin{eqnarray}
\label{5.23}
f_- &=& G_R \Delta^-_{q_-,\pi}\bar{G}_R + \ldots \\
\label{5.24}
f_+ &=& G_R\Delta^+_{q_-,\pi}\bar{G}_R + G_R\Delta^+_{q_+,\pi}\bar{G}_R
+ \cdots\>,
\end{eqnarray}
where $\Delta^-_{q_-,\pi}(p)$ and $\Delta^+_{q_-,\pi}(p)$, 
\begin{eqnarray}
\label{5.25}
\Delta^-_{q_-,\pi}(p) &=& g^2 \int_{\left(\mu+\sqrt{p^2}\right)^2}^{\infty} d\kappa^2
\left[\, a_-(\kappa^2)I_1(p,\kappa,\mu) + b_-(\kappa^2)I_2(p,\kappa,\mu)\,\right] +
\ldots  , \\
\label{5.26}
\Delta^+_{q_-,\pi}(p) &=& g^2 \vartheta(\mu-\sqrt{p^2})
 \nonumber\\
 && \times  \int_0^{\left(\mu-\sqrt{p^2}\right)^2} d\kappa^2
\left[\, a_-(\kappa^2)I_1(p,\kappa,\mu) + b_-(\kappa^2)I_2(p,\kappa,\mu)\,\right] + \ldots,
\end{eqnarray}
are the contributions corresponding to the quark masses $\kappa$ larger
($\kappa>\mu$) and smaller ($\kappa<\mu$) than the $\pi$-meson mass
$\mu$, respectively. 
Higher order terms have the same structure:
large quark masses of $f_-$ contribute to $f_-$, small quark masses contribute
to $f_+$. 

The result of this analysis is the following. The imaginary
part of $f_-(p^2)$ which equals zero in a non-confined theory has to
satisfy a non-linear homogeneous equation. Due to asymptotic freedom,
it must decrease at large $p^2$. In linear approximation the equation
for $f_-$ with zero boundary conditions at infinity is a typical
bound state equation defining the quark distribution in the vacuum.

The equation for $f_+$ is very different. The solution of the equation
for $f_-$ gives an inhomogeneous term for the equation for $f_+$.
The latter becomes similar to the equation for $f_+$ in a non-confined theory
(\ref{5.13}). 
Now, instead of the usual $f_0(p) = \delta(p^2-m^2)(\hat{p}+m)$,
we have 
\begin{equation}
\label{5.27}
f_0(p) \>=\> G_R\, \Delta_{q_-,\pi}^+(p)\, \bar{G}_R\>.
\end{equation}
Note that, according to (\ref{5.26}), $f_0(p)$ 
is proportional to $\vartheta(\mu-\sqrt{p^2})$. 
Iterating,
we have to find $f_+$ which corresponds to perturbation theory
at large $p^2$.

The equations for $f_+$ and $f_-$ contain $G_R$. If we were able
to find $f_+$, $f_-$ for a given $G_R$, we would then obtain an
equation for $G_R$ in the form of the dispersion relation
\begin{equation}
\label{5.28}
G_R(p) \>=\> \frac{1}{\pi} \int \frac{dp'_0}{p'_0-p_0} f(p'_0,p) -
\frac{1}{\pi} \int \frac{dp'_0}{p'_0-p_0} \tilde{f}(p'_0,p)\>.
\end{equation}
This $G_R(p)$ has no poles since there are no $\delta$-functions
in $f_+$ and $f_-$.
The equations (\ref{5.16}), (\ref{5.17}) can be considered as
generalised unitarity relations. 

The structure of the equations for
gluonic $f_{g+}$, $f_{g-}$ is different only in the sense that
$f_{g-}$ 
contains an inhomogeneous term corresponding to intermediate quark
states.

All these equations are in excellent agreement with the idea that
when the negative energy states are not completely occupied, both
positive and negative energy states become unstable. 
The decay modes of these states are explicitly written in the equations
(\ref{5.16}), (\ref{5.17}).

Up to now we ignored the contribution $f_{Eucl}(p)\vartheta(-p^2)$. 
Let us include it in the right-hand sides of
(\ref{5.16}), (\ref{5.17}). The equation for $f_-$ will remain the same
homogeneous equation 
(since it is impossible to have a time-like
vector $p$ with a negative $p_0$ as a sum of a space-like
quark momentum $p'$ and a time-like pion momentum $k$ with a positive $k_0$).
The equation for $f_+$ (\ref{5.24}) will acquire an additional term:
\begin{equation}
\label{5.29}
f_+ \>=\> G_R\,\Delta^+_{q_-,\pi}\,\bar{G}_R 
+ G_R\,\Delta^+_{q_+,\pi}\,\bar{G}_R +
G_R\,\Delta^+_{q_{Eucl},\pi}\,\bar{G}_R\>.
\end{equation}
The equation for $f_{Eucl}$ can be written in the form
\begin{equation}
\label{5.30}
f_{Eucl}(p) \>=\> G_R\,\Delta^{Eucl}_{q_-,\pi}\,\bar{G}_R +
G_R\,\Delta^{Eucl}_{q_+,\pi}\, \bar{G}_R \>,
\end{equation}
containing a driving term. 
The iterative solution of (\ref{5.30}) has the following  
diagrammatic structure: 
\begin{equation}
\label{5.31}
\begin{picture}(0,0)%
\epsfig{file=fig52.pstex}%
\end{picture}%
\setlength{\unitlength}{4144sp}%
\begingroup\makeatletter\ifx\SetFigFont\undefined%
\gdef\SetFigFont#1#2#3#4#5{%
  \reset@font\fontsize{#1}{#2pt}%
  \fontfamily{#3}\fontseries{#4}\fontshape{#5}%
  \selectfont}%
\fi\endgroup%
\begin{picture}(924,493)(1564,-2312)
\put(1846,-1951){\makebox(0,0)[lb]{\smash{\SetFigFont{12}{14.4}{\rmdefault}{\mddefault}{\updefault}$f_{Eucl}(p)$}}}
\end{picture}
\,=\, \begin{picture}(0,0)%
\epsfig{file=fig53.pstex}%
\end{picture}%
\setlength{\unitlength}{4144sp}%
\begingroup\makeatletter\ifx\SetFigFont\undefined%
\gdef\SetFigFont#1#2#3#4#5{%
  \reset@font\fontsize{#1}{#2pt}%
  \fontfamily{#3}\fontseries{#4}\fontshape{#5}%
  \selectfont}%
\fi\endgroup%
\begin{picture}(1194,844)(1429,-2495)
\put(2206,-2446){\makebox(0,0)[lb]{\smash{\SetFigFont{12}{14.4}{\rmdefault}{\mddefault}{\updefault}$\pi$}}}
\put(2161,-1816){\makebox(0,0)[lb]{\smash{\SetFigFont{12}{14.4}{\rmdefault}{\mddefault}{\updefault}$f_-$}}}
\end{picture}
\,
+\, \begin{picture}(0,0)%
\epsfig{file=fig54.pstex}%
\end{picture}%
\setlength{\unitlength}{4144sp}%
\begingroup\makeatletter\ifx\SetFigFont\undefined%
\gdef\SetFigFont#1#2#3#4#5{%
  \reset@font\fontsize{#1}{#2pt}%
  \fontfamily{#3}\fontseries{#4}\fontshape{#5}%
  \selectfont}%
\fi\endgroup%
\begin{picture}(1194,796)(529,-647)
\put(1261,-16){\makebox(0,0)[lb]{\smash{\SetFigFont{12}{14.4}{\rmdefault}{\mddefault}{\updefault}$f_-$}}}
\end{picture}
\,+\, \>\> \ldots \hspace{.4cm} .
\end{equation}
Inserting this solution into the last term of (\ref{5.29}) we will
have an equation similar to (\ref{5.24}) with the only difference that
$\Delta^+_{q_-,\pi}$ in (\ref{5.24}) acquires an additional contribution
from multi-pion states. 
This will not change any of the conclusions we have reached, because 
it is only  $f_+$ and $f_-$ which matter 
for the calculation of the retarded Green's function.

The structure of $f_{Eucl}$ written in (\ref{5.31}) is, nevertheless,
interesting in two respects. 
Firstly, it shows 
how 
the analytic features corresponding to Fig.~\ref{Fig17} 
(with functions having discontinuities at positive and negative $p^2$ values)
can be reduced to analytic features corresponding to Fig.~\ref{Fig18} where
the discontinuity differs from zero only at positive $p^2$ values.
Secondly, the existence of the integrals 
in 
(\ref{5.31}) imposes a restriction on $f_-(p)$.

\section*{References}
\def\labelenumi{[\arabic{enumi}]}

\begin{enumerate}
\item V.\,N.\ Gribov, Physica Scripta {\bf T 15} (1987), 164
\item Vladimir N.\ Gribov, Orsay lectures on confinement (II), 
LPTHE Orsay 94-20 (1994), hep-ph/9404332
\item V.\,N.\ Gribov, Lund preprint LU-TP 91-7 (1991)
\item F.\,E.\ Close, Yu.\,L.\ Dokshitzer, V.\,N.\ Gribov, V.\,A.\ Khoze, 
M.\,G.\ Ryskin, Phys.\ Lett.\ {\bf B 319} (1993), 291
\item V.\,N.\ Gribov, 
{\sl Bound states of massless fermions as a source for new physics}, 
in Proceedings of the International School of Subnuclear Physics, 
33th Course, Erice, Italy, 1995, ed.\ A.\ Zichichi, World Scientific, p.\ 1, 
Bonn preprint TK-95-35 (1995), hep-ph/9512352
\item V.\,N.\ Gribov, Bonn preprint TK-97-08 (1997), 
hep-ph/9807224 (annotated version)
\item V.\,N.\ Gribov, 
{\sl The theory of quark confinement}, 
in Proceedings of the International School of Subnuclear Physics, 
34th Course, Erice, Italy, 1996, ed.\ A.\ Zichichi, World Scientific, p.\ 30
\end{enumerate}
\end{document}